# Multi-output neural network surrogate for double- pulse femtosecond optothermal control of thin silicon films on fused silica

GEORGE D. TSIBIDIS[1*]

[1]Institute of Electronic Structure and Laser (IESL), Foundation for Research and Technology (FORTH), Vassilika Vouton, 70013, Heraklion, Crete, Greece

* Email: tsibidis@iesl.forth.gr

## Abstract

**We introduce a multi-output feedforward neural network surrogate trained on a coupled multiple-reflection and two-temperature model to map double-pulse femtosecond laser excitation (1030 nm) of crystalline silicon films on fused silica across film thicknesses (50-500 nm) and inter-pulse delays (0-10ps). Trained via Levenberg-Marquardt optimization on 2000 physical simulations, the dual-hidden-layer network ([40,30] neurons) simultaneously predicts six strongly coupled optothermal observables, peak electron temperature, peak lattice temperature, peak carrier density, minimum and maximum absorptivity and minimum refractive index achieving coefficients of determination $R^2 \geq 0.9998$ across all targets. Evaluating a dense 200×200 design grid in milliseconds resolves pronounced thin-film interference fringes and carrier-induced index modulation, identifying an optimal optothermal coupling regime at $d$=156 nm and $\Delta\tau$=4.572 ps. This unified machine-learning workflow eliminates computationally expensive multi-physics runtime bottlenecks for real-time laser processing control and inverse photonic thin-film design.**

Femtosecond laser processing of semiconductor thin films on transparent dielectrics enables selective ablation, crystallization and nanophotonic structuring [1-3]. At a near-infrared wavelength of 1030 nm (i.e. where photon energy is above the indirect band gap of crystalline silicon), single photon inter-band absorption in sub-micrometer silicon films is weak; thus, laser-matter coupling is dictated by nonlinear two-photon/free-carrier absorption and thin-film interference [4]. As the dense photoexcited electron-hole plasma dynamically reduces the real part of the refractive index during and immediately after excitation [5], a temporarily delayed second pulse 'sees' an optically modified material whose transient reflectivity and absorptivity diverge significantly [6]. Tailoring the inter-pulse $\Delta\tau$ alongside the film thickness $d$ therefore provides deterministic control over localized energy deposition [7].

On the other hand, exploring this coupled opto-thermal parameter space via direct numerical integration is computationally expensive. Resolving self-consistent wave propagation alongside non-equilibrium carrier and lattice transport requires solving stiff partial differential equations at sub-femtosecond optical time steps for every geometry and pulse sequence. While machine-learning surrogates and inverse-design frameworks have accelerated static nanophotonic optimization [8], ultrafast laser-semiconductor interactions introduce a major challenge: optical and thermal observables are strongly coupled across various scales, spanning four orders of magnitude in lattice temperature and carrier density while exhibiting narrow, periodic Fabry-Pérot interference resonances in thickness [9, 10]. Training isolated single-output regressors often breaks physical consistency between extrema of refractive index, absorptivity and thermal peak states. Here, we demonstrate a compact multi-output artificial neural network (ANN) surrogate [11] that learns all six primary optical and thermal observables from a 40×50 thickness-delay simulation grid, enabling rapid high-resolution mapping and multi-objective optothermal optimization.

The physical training dataset is generated by coupling a one-dimensional transfer-matrix multiple reflection solver at normal incidence ($\lambda_L$= 1030 nm) with a semiconductor two-temperature model (TTM) for a crystalline silicon film of thickness $d$ on a semi-infinite fused-silica substrate (of refractive index $n_s$=1.45 [12]). The sample is irradiated by two identical Gaussian laser pulses, each with a pulse duration of 170 fs and a peak fluence of 0.2 J/cm², separated by a time delay of $\Delta\tau$ (i.e. the laser parameters considered in this study were selected to serve as a proof of concept; however, the proposed approach is general and can be extended to a wide range of laser conditions). The silicon layer is discretized into spatial sub-layers matching the thermal mesh where forward- and backward propagating electric fields yield transient reflectivity $R$, transmissivity $T$ and absorptivity $A$=1-$R$-$T$. The transient complex permittivity of each sub-layer undergoes dynamic evolution as the laser beam, with intensity $I$, excites electrons, resulting in a temporarily varying carrier density $N_e$ provided by [4, 10, 13]

$$\frac{dN_e}{dt} = \frac{\alpha_{SPA}}{\hbar\omega_L} I + \frac{\beta_{TPA}}{2\hbar\omega_L} I^2 - \gamma (N_e)^3 + \theta N_e - \vec{\nabla} \cdot \vec{J} \qquad (1)$$

In Eq.1 $\alpha_{SPA}$, $\beta_{TPA}$, $\gamma$ stand for the one-photon, two-photon and Auger recombination, respectively, $\vec{J}$ is the carrier current density [4, 10, 13]; $N_e$ corresponds to the carrier density inside Si while $\theta$ stands for the impact ionisation coefficient[4, 10, 13]. The coupled thermal response of the electron and lattice subsystems is governed by the two-temperature model [4, 10, 13]

$$\begin{aligned} C_e \frac{\partial T_e}{\partial t} &= \frac{\partial}{\partial z}\left(k_e \frac{\partial}{\partial z} T_e\right) - G_{cL}(T_e - T_L) + S \\ C_L \frac{\partial T_L}{\partial t} &= \frac{\partial}{\partial z}\left(k_L \frac{\partial}{\partial z} T_L\right) + G_{cL}(T_e - T_L) \end{aligned} \qquad (2)$$

where the electron $T_e$ and lattice $T_L$ temperatures, respectively, of Si are calculated while $C_e$ and $C_L$ stand for the heat capacities of the carriers and the lattice, respectively, $G_{cL}$ is the carrier-phonon coupling, $k_L$, ($k_e$) correspond to the lattice's (carrier's) thermal conductivity and $S$ stands for a generalized 'source' term given by the following expression [14, 15].

Results for the dynamics of the intensity profile, refractive index, absorptivity, $N_e$, $T_e$, $T_L$ in Si are illustrated in Figure 1 considering two temporarily separated pulses ($\Delta\tau$=3 ps). As shown in Figure 1a, the absorptivity exhibits a distinct two-step jump rising to ~0.036 after the first pulse while it reaches ~0.082 after the second pulse dictated by nonlinear absorption. This behaviour originates from the internal optical intensity profile (Figure 1b), which establishes periodic standing-wave antinodes (i.e. at $z\approx 55\,nm$ and $z\approx$190 nm) across the film depth during each pulse [10]. As a result, the generated electron-hole plasma reduces the local refractive index $n$ from its pristine value of 3.5632+$i$0.00027806 at $\lambda_L$=1030 nm [10] to ~3.42 nm at the antinodes as shown in Figure 1c. This corresponds to the carrier density increase across the delay window in Figure 1d which introduces a modified refractive index profile for the incoming second pulse. The reduced refractive index of the excited solid facilitates the enhancement of energy absorption during the second pulse, generating peak electron temperatures above 20×10$^3$ K localized at the standing-wave antinodes as depicted in Figure 1e. The electron excitation is followed by sub-picosecond electron-phonon relaxation that induces localized lattice thermal hotspots $T_L$~530 K) at the subsurface and buried interface antinodes (Figure 1f).

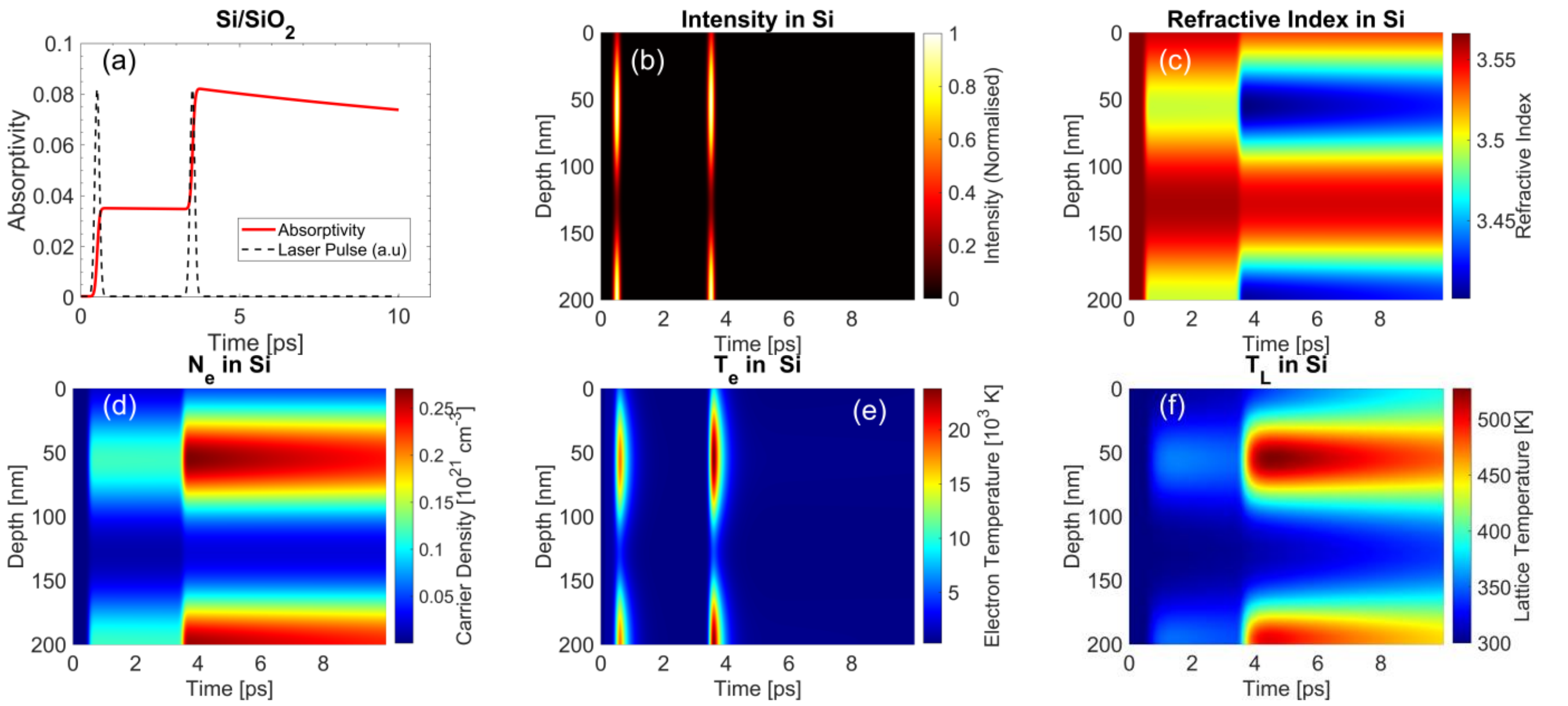

**Figure 1:** Spatiotemporal dynamics of a 200 nm $Si/SiO_2$ thin film under double-pulse irradiation of $\Delta\tau$=3 ps: (a) Absorptivity, (b) Internal optical intensity, (c) Transient refractive index, (d) Carrier density, (e) Electron temperature, and (f) Lattice temperature profiles.

Since this interplay couples the optical, carrier and thermal responses non-monotonically across thickness and delay, mapping the full ($d$, $\Delta\tau$) profile by extensive simulations is computationally expensive. We, therefore, develop a multi-output neural-network surrogate of the coupled optics-carrier-thermal model which enables efficient exploration of the full thickness-delay parameter space. To achieve this, we sample a uniform 40×50 design grid spanning $d \in [50,500]$ nm and $\Delta\tau \in [0,10]$ ps, yielding $N$=2000 multiphysics simulations. For each run, scalar response metrics are extracted: max($T_e$), max($T_L$), max($N_e$), min($A$), max($A$), minimum transient refractive index (min($n$ )). To develop a fast, precise, unified surrogate capable of preserving cross-correlations between optical modulation and thermal response, we implement a multi-output feedforward neural network architecture rather than independent single-variable Gaussian-process regressors [16]. Both the two-dimensional input vectors $\mathbf{x}=[d, \Delta\tau]^T$ and the six-dimensional target vectors are linearly normalized to the [-1, 1] interval via min-max scaling prior to training so that a broad range of scales ranging from min($A$)~$10^{-3}$ to max($N_e$)~$10^{21}$ $cm^{-3}$ contribute balanced gradients during backpropagation. The network comprises two fully connected hidden layers with 40 and 30 nonlinear neurons, respectively, mapping into a six-node linear output layer. The 2000 samples are randomly partitioned into 70% training (1400 runs), 15% validation (300 runs) and 15% independent testing (300 runs) sets and network weights are optimized using the second-order Levenberg-Marquardt backpropagation algorithm with early stopping monitored on the validation subset [17]. Figure 2 demonstrates the regression accuracy of the trained multi-output surrogate across all physical observables. Every subplot shows tight clustering along the $y_{predicted}=y_{true}$ parity diagonal across the entire dynamic range of the simulation dataset, confirming that the shared hidden representation simultaneously exhibit pronounced interference extrema and nonlinear plasma thresholds without overfitting. More specifically, the optical metrics achieve near-exact reconstruction, with $R^2 \geq 0.9998$ for all six parameters. Once trained, the surrogate enables sub-millisecond predictions, allowing surrogate-based interpolation to a 200×200 grid (40,000 parameter combinations). Figure 3 illustrates the resulting continuous multi-physical response surfaces across $d$ and $\Delta\tau$. The maps reveal pronounced periodic vertical stripes dictated by the thin-film Fabry-Pérot interference inside the silicon layer, with a fundamental half-wave thickness periodicity of $\lambda_L/(2n) \approx 144$ nm. At constructive internal field resonances near $d \approx 145$-155 nm, $d$≈295-305 nm and $d$≈445-465 nm, an enhanced (intra-cavity) intensity drives carrier generation to up to 3.7×$10^{21}$ cm$^{-3}$ and $T_e$ above 6.5×$10^4$ K. Interestingly, the refractive index drops from its near pristine value to ~1.62 inside the resonance band around $d$~145 nm.

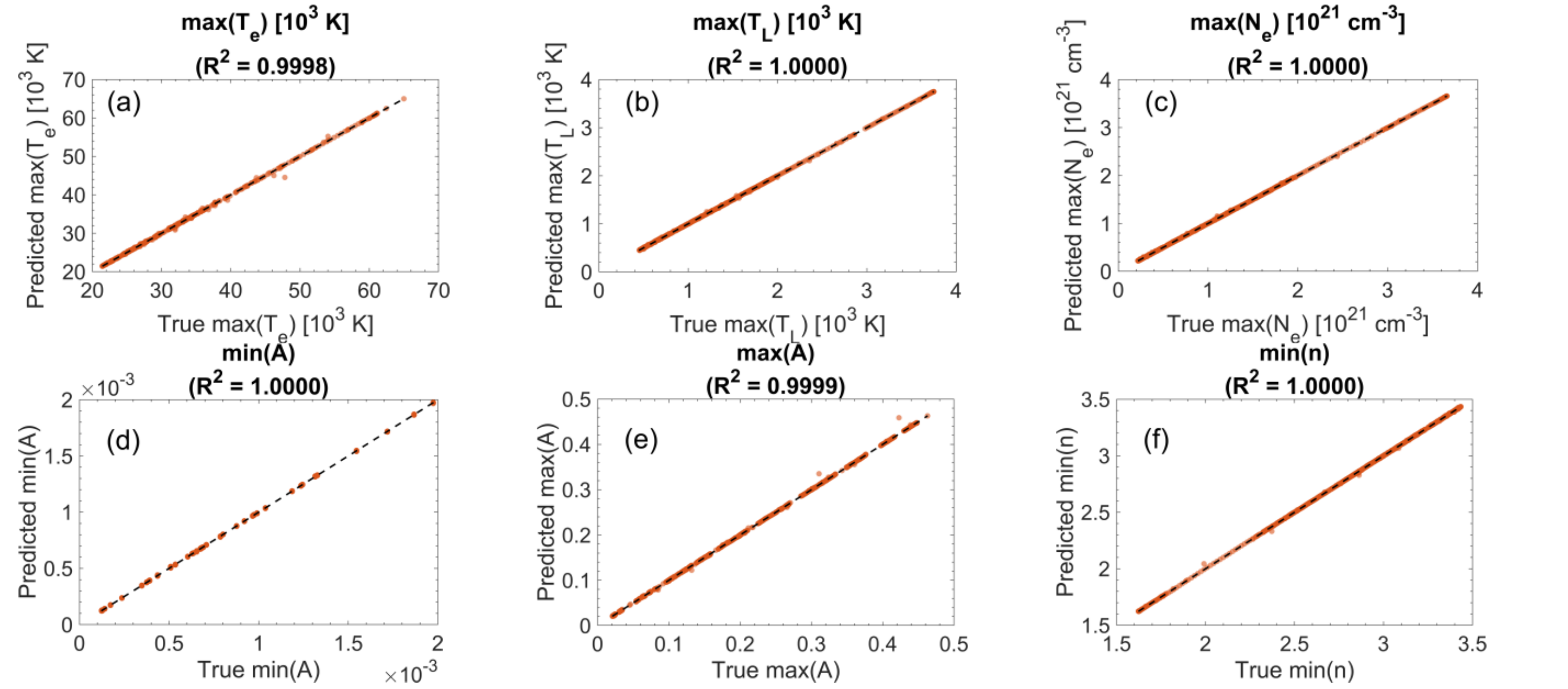

**Figure 2:** Multi-output neural network surrogate parity validation across all coupled multiple-reflection/TTM simulations for the six target metrics: max($T_e$), max($T_L$), max($N_e$), min($A$), max($A$), and min($n$). *Dashed* black lines denote ideal unity parity ($y_{predicted}$=$y_{true}$) that is confirmed from $R^2 \geq 0.9998$.

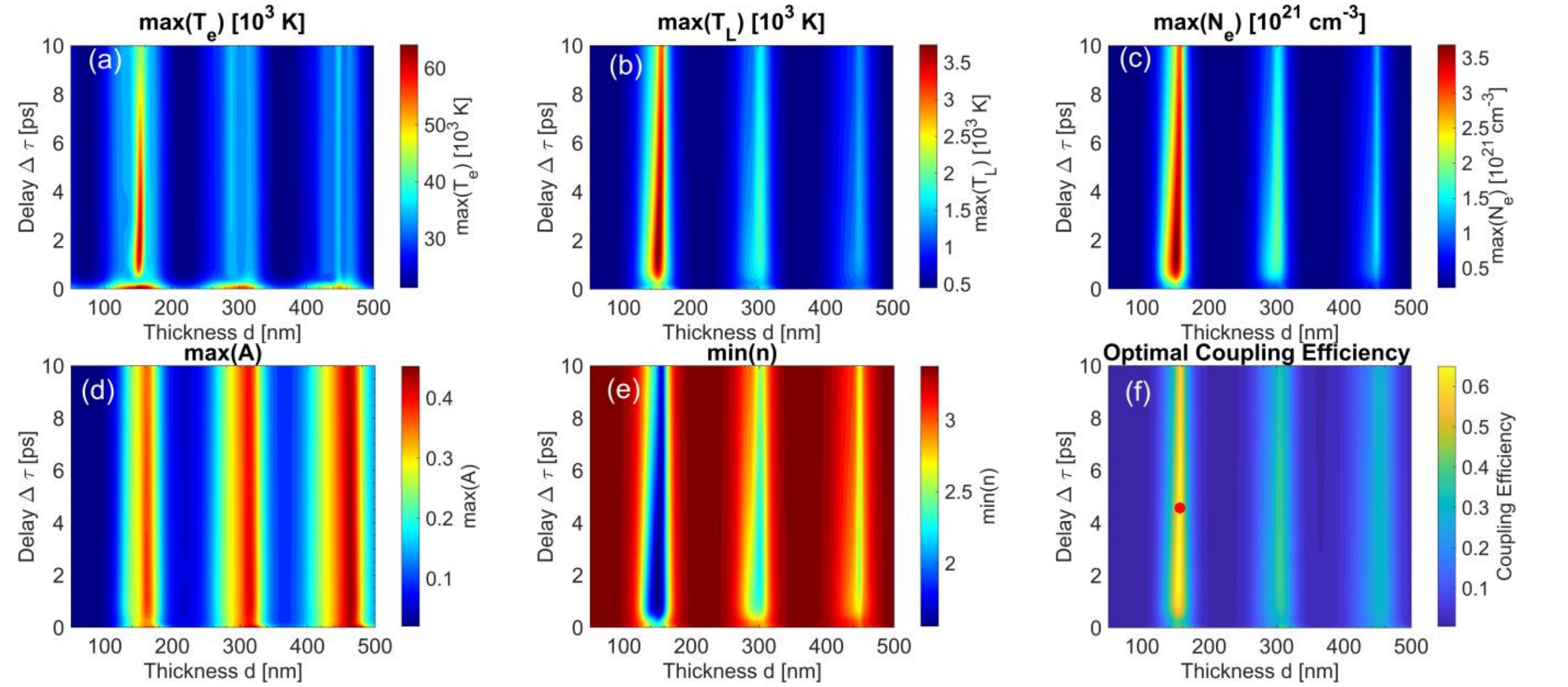

**Figure 3:** Surrogate-predicted parameter maps versus $d$ and $\Delta\tau$: (a) max($T_e$), (b) max($T_L$), (c) max($N_e$), (d) max($A$), (e) min($n$) and (f) multi-objective optimal coupling efficiency with the global optothermal optimum highlighted by the *red* marker.

A comparison of Figures 3a,b highlights the competing effects of film thickness and pulse delay on double-pulse laser processing. Maximum absorption (~0.46) occurs in thicker films (~465 nm), due to longer optical path lengths, whereas maximum lattice heating (>~3800 K) is localized around 150 nm where constructive interference constrains the absorbed energy within a smaller thermal volume, exceeding silicon's melting point (1687 K). Both thermal and carrier responses also exhibit a non-monotonic dependence on pulse delay with enhanced coupling at ~0.5-4 ps. This enhancement arises from free-carrier absorption induced by the plasma generated by the first pulse while carrier diffusion and Auger recombination progressively reduce excitation at longer delays.

To demonstrate how the AI surrogate enables rapid inverse process design, a normalized multi-objective optothermal coupling efficiency metric *EM* is defined, ($\equiv \max(T_L)/\max(T_L)_{global} \times \max(A/\max(A)_{global})$ evaluated directly across all 40,000 surrogate grid points (Fig. 3f). This complex efficiency profile reveals a well-defined optothermal optimum at ($d$, $\Delta\tau$)=(156 nm, 4.57ps) (*red marker*, where *EM*~0.66). This identifies the specific combination of film thickness and pulse delay that provides the most favourable balance between optical absorption and localized lattice heating. The identification of this narrow resonance peak via rigorous multi-physics investigation would require tens of thousands of computationally expensive TTM-Maxwell runs, whereas the multi-output neural network yields the value in milliseconds.

The above results provide a basis for modelling optical media whose properties evolve during laser-matter interaction. In laser nanomanufacturing, pump-probe spectroscopy and reconfigurable phase-change photonics, carrier-induced index changes undermine static thin-film design rules. A precise surrogate of the coupled optics-carrier model enables rapid inversion, uncertainty propagation for the nonlinear material coefficients (e.g., absorption coefficients, impact-ionization rate, electron collision time [13, 14]), and control of pulse separation and film thickness. The surrogate inherits the approximations of the underlying solver (one-dimensional propagation, plane-wave illumination, no phase transformations) and is valid only within the sampled range, $d \in [50, 500]$ nm and $\Delta\tau \in [0, 10]$ ps. The same framework is capable, however, to be extended to higher-dimensional design spaces that include fluence, pulse duration, wavelength and multi-pulse trains.

In summary, we have presented a neural-network surrogate for 1030 nm double-pulse femtosecond excitation of silicon films on fused silica. Through a training on 2000 coupled multiple-reflection and two-temperature simulations, it reproduces six optical, electronic and thermal values with $R^2 \geq 0.9998$ and evaluates 40,000-point multi-physics maps in milliseconds. These maps resolve the interplay of Fabry-Pérot thickness resonances and refractive index modulation and identify an optimal coupling regime at $d = 156$ nm and $\Delta\tau = 4.57$ ps. Coupled to damage and ablation criteria [18], such surrogates could predict damage thresholds as a function of film thickness, pulse delay and fluence. This would enable the efficient design of experimental configurations and manufacturing protocols [19] and reduce the experimental effort needed to identify suitable processing windows.

**Funding.**This research work was supported from Horizon Europe, the European Union's Framework Programme for Research and Innovation, under Grant Agreement No. 101057457 (METAMORPHA).

**Data availability.** Data underlying the results may be obtained from the authors upon reasonable request.

**Author Disclosures.** The authors declare no conflicts of interest.